\documentclass[physics,article,accept,pdftex,moreauthors]{Definitions/mdpi} 

\firstpage{94} 
\pubvolume{6}
\issuenum{1}
\articlenumber{0}
\pubyear{2024}
\copyrightyear{2024}
\datereceived{5 October 2023} 
\daterevised{1 December 2023} 
\dateaccepted{5 December 2023} 
\datepublished{10 January 2024} 
\hreflink{https://doi.org/10.3390/\linebreak physics6010007}
\doinum{10.3390/physics6010007}

\usepackage{amsmath, amsthm, amssymb, amsfonts}

\Title{Zero-Point Energy Density at the Origin of the Vacuum Permittivity and Photon Propagation Time Fluctuation}

\TitleCitation{Zero-Point Energy Density at the Origin of the Vacuum Permittivity and Photon Propagation Time Fluctuation}

\Author{{Christophe Hugon\orcidA{}} 
 $^{1}$
and Vladimir {Kulikovskiy\orcidB{}}
 $^{2}$}

\AuthorNames{Christophe Hugon and Vladimir Kulikovskiy}

\AuthorCitation{Hugon, C.; Kulikovskiy, V.}

\address{%
$^{1}$ \quad {R}\&{DoM} (Research \& Development on Marseilles), 65 Bd de la Lib\'{e}ration, 13001 Marseille, France; chr.hugon@protonmail.com\\
$^{2}$ \quad Istituto Nazionale di Fisica Nucleare, {INFN Sezione di Genova}, Via Dodecaneso 33, 16146 Genova, Italy; vladimir.kulikovskiy@ge.infn.it}

\abstract{We give a vacuum description with zero-point density for virtual fluctuations. One of the goals is to explain the origin of the vacuum permittivity and permeability and to calculate their values. In particular, we improve on existing calculations by avoiding assumptions on the volume occupied by virtual fluctuations. We propose testing of the models that assume a finite lifetime of virtual fluctuation. If during its propagation, the photon is stochastically trapped and released by virtual pairs, the propagation velocity may fluctuate. The propagation time fluctuation is estimated for several existing models. The obtained values are measurable with  available technologies involving ultra-short laser pulses, and some of the models are already in conflict with the existing astronomical observations. The phase velocity is not affected significantly, which is consistent with the interferometric measurements.}


\begin{document}

\section{Introduction}
\label{s:introduction}
With the concept of virtual fluctuations composed of two photons, several effects can be introduced and numerically estimated. This includes the known Lamb shift measured in the Lamb–Retherford experiment, the measured Casimir effect~\cite{Lamoreaux1997}, the observed dynamic Casimir effect~\cite{Wilson2011}, and the predicted Unruh effect, as well as Hawking radiation. There is no general consensus, however, whether the virtual fluctuations are 
rather a physical phenomenon than just a useful mathematical tool~\cite{RefVirtPart}.

Quantum mechanics (QM) is based on  postulated equations which do not have an intuitive introduction despite and because of the more than a dozen quite contradictory interpretations available~\cite{Cavalleri2009, Cetto2022b}. The attempts to derive some of the principles of QM through classical stochastic processes are ongoing in order to provide a deeper understanding of the (experimental) phenomena from a wider point of view and context. In particular, in  stochastic electrodynamics (SED), the interaction of the zero-point field (ZPF) with real particles is evaluated. This interaction may explain several, if not all, of the quantum phenomena (for one of the most recent 
papers; see \cite{Cavalleri2009}). The energy density of the ZPF, $w(\omega)\propto\omega^3$, 
 can be derived from the condition that there is no average force of 
the ZPF acting on any physical harmonic oscillator with a frequency $\omega$ (Einstein--Hopf formula~\cite{Boyer1969} (Appendix)). 

Thus, this energy density has the same form as ZPF in 
quantum electrodynamics (QED). Some of the 
quantum relations, such as the canonical commutation relation, $[\hat x; \hat p] = i\hbar${, with $\hat x$ and $\hat p$ the coordinate and momentum operators, respectively, and $\hbar$ the reduced Planck's constant, 
$h/(2\pi)$,} 
can be obtained in SED only if the interaction with ZPF is treated in a non-perturbative way~\cite{Cetto2022b,Huang2020}. This  is true in particular for free particles: the transition from a classical deterministic behaviour to an indeterministic quantum one happens in SED once the interaction with ZPF takes the leading role and the information on the initial condition is lost~\cite{Cetto2022a}.

In QED, the virtual fluctuations or ZPF manifest themselves as an additional $1/2\hbar\omega$ term of the total energy stored in a single oscillation mode. 
That term appears for each of the light modes for the photons in the box, and the number of modes becomes continuous once the infinite box size limit is 
considered for  Planck's law derivation~\cite{Planck1912}. Conventionally, the $1/2\hbar\omega$ term is omitted in order to evaluate the energy difference {with} respect to the so-called zero-point level. This 
is sufficient for  most applications, where only the energy difference matters. Nevertheless, the zero-point energy is not null even in the absence of real particles 
(photons), and it affects gravity at the cosmological scales. Moreover, the mode's energy density is 
$w(\varepsilon)\propto\varepsilon^3$, 
infinite for $\varepsilon\to\infty$.  In the framework of quantum field theory (QFT), 
generally, the energy upper limit at the Planck scale, $\Lambda$, is hypothesized. As a result, the zero-point energy density $w\propto\Lambda^4$ has up to 120 orders of difference with the observed energy density---an issue known as the cosmological constant problem.

In the present study, 
the concept of virtual fluctuations composed by virtual fermion--antifermion pairs is explored. The motivations to enrich virtual fluctuations with fermion--antifermion pairs are not new, and overviews can be found in
 {Refs.}~\cite{RefUrban, Mainland2020, Leuchs2023}. Here, we use virtual fluctuations as a synonym for virtual pairs and mostly consider only fermions and antifermions. 

In Refs.\cite{RefUrban,Leuchs2023}, 
it is assumed that the virtual pairs may appear for a short lifetime connected with their energy by the Heisenberg uncertainty relation. The pairs are CP-symmetric, 
what permits them to have zero values for the total angular momentum, colour, and spin. In the presence of the electromagnetic field, the pairs should polarise, and thus the vacuum has dielectric behaviour. The vacuum's dielectric properties, i.e., 
 {the permittivity,} 
$\epsilon_0$, and the permeability, 
$\mu_0$, are already axiomatically postulated in  Maxwell's equations. One can, however, take a step back and reconsider this 
by looking at the following equations of the {electric} displacement field, $\vec{D}$, 
and the {magnetic field strength,} $\vec{H}$,  for dielectrics:
\begin{align}
\vec{D}=\epsilon_0 \vec{E} + \vec{P},
\label{eq1}
\\
\vec{H}=\frac{1}{\mu_0} \vec{B} - \vec{M}, 
\label{eq2} 
\end{align}
 with $\vec{E}$ and  $\vec{B}$ being {electric field strength and magnetic flux density}, respectively. 

Polarisation, $P$, and magnetisation, $M$, induced by the external field at the microscopic level correspond to the sum of electric and magnetic moments in a unit of volume, respectively. Following
Refs.~\cite{RefUrban,Leuchs2023}, 
one can consider that the first terms  in Equations (\ref{eq1}) and (\ref{eq2}) are due to the vacuum polarisation and magnetisation, i.e., these terms can also be estimated as the sums of electric and magnetic moments of the virtual fermion pairs.

In order to calculate $\epsilon_0$ and  {$1/\mu_0$,} 
the moments generated by a virtual pair are needed, along with the volume occupied by each pair. For the moments' estimation, the calculations in Ref. \cite{Leuchs2023,Margan2011,Mainland2020} take a shortcut by assuming 
the oscillation model for the virtual pair with two states separated by $2 mc^2$ with $m$ the particle mass and $c$ the speed light in the vacuum.  The calculations in Ref. \cite{RefUrban} 
start from the dipole moment of a pair with opposite charges and  fermion magnetic moment and then assume that the pair lifetime is modified in the presence 
of the field. For the volume estimation, the typical size of a Compton length is commonly involved. In particular,  in~\cite{Leuchs2023}, this is 
motivated by the assumption that in order for the virtual pair to interact with the external field, the energy conservation should be violated locally by 
$\Delta \varepsilon \gtrsim 2 mc^2$, which is non-detectable in a period shorter than $\hbar/(2 mc^2)$. 
If the speed of particles is at the maximum, i.e., at the speed of light, $c$, the pair must remain separated by a distance 
$\lambda_C = \hbar/mc$. 
In order to obtain a measured value of $\epsilon_0$, the average volume of $V\simeq0.41 \lambda_C^3$ is required, and it is equally occupied by the virtual fermions of all known types.

In the study presented here, we address 
the assumptions on the volume occupied by virtual fluctuations in the above mentioned models. We consider that a self-consistent and intuitive way to introduce the occupied volume is to go back to the assumed origin of virtual fluctuations, i.e., ZPF which appears in solutions of QED equations or which is introduced from the beginning in SED. One already knows the ZPF energy density and the 
energy per mode. Thus, the ZPF itself already provides the modes' density, $\rho(\varepsilon)$. The mode's density can be used in order to obtain 
any property density if the expression of the property per mode is known. The assumption of the infinity of the modes and a distinct energy associated with each mode is incompatible with the assumption that virtual fluctuations appear only with total energy, $\varepsilon 
= 2 mc^2$. 
Another feature is that by introducing ZPF density, one does not need  to further assume that the virtual pairs become real for the time related to 
their energy following the Heisenberg uncertainty. The only assumption needed is that at the ground state the energy of virtual fluctuation is
$\varepsilon=1/2\hbar\omega$, while after interaction, it becomes $\varepsilon=(1/2+n)\hbar\omega$. In QED, this means that before the interaction, no real particles were present, while after the interaction, $n$ particles were produced. In SED, the virtual fluctuations are  part of the ZPF, which is real even before or in the absence of \mbox{an interaction.}

In Section~\ref{sec:density}, we review the Planck's law derivation in order to introduce the density of virtual fermions  and to explain their mathematical origin. We show how $\epsilon_0$ can be estimated using vacuum density in Section~\ref{sec:MuEpsi}.

One may further assume that the photon propagation speed is actually finite only because the photon is delayed at each interaction with virtual fluctuations by their annihilation time. The real photon propagation  then fluctuates, without affecting the phase velocity measured by interferometers~\cite{UrbanReply}, and the arrival time spread can be estimated by knowing the average lifetime of a virtual pair and the distance travelled. This is performed for several theories in Section~\ref{sec:prop}, and such spread presents a new prediction, measurable with the available technologies.

\section{Statistics for Virtual Pairs}
\label{sec:density}

The conventional derivation of the Planck's law is briefly summarised in this Section with the explicit assumptions needed. In many textbooks, these calculations are oriented to provide a measurable energy density, so  the vacuum-related terms are often hidden and the physical origins behind some assumptions are not reported properly. 

\subsection{Density of Modes for a Particle in a Box}
\label{s:density_of_modes}
\textls[-52]{A particle in a box may only have modes respecting the boundary conditions} \mbox{$\psi (0)=\psi (L)=0$}, where $L$ is the size of a box.
\textls[-25]{This is satisfied for standing waves which have half-wavelength {($\lambda/2$)}
multiplets equal to a box size}:
\begin{equation}
 l_i\frac{\lambda_i}{2}=L_i,
 \end{equation}
 for every space coordinate $i$.  This condition shows that states are discrete for a box with a limited size. Each set of non negative integer numbers ${l_i}$ defines a mode.
  The  wavelength is connected with \mbox{momentum $p_i$ as}:
 \begin{equation}
 \lambda_i=h/p_i\, .
 \end{equation}

So, the number of states in the mode space corresponds to the number of the states in the momentum space as follows:
\begin{equation}
dl_xdl_ydl_z=\frac{8L_xL_yL_z}{h^3}dp_xdp_ydp_z=\frac{8V}{h^3}d^3p,
\end{equation}
for a box with volume $V$.

For a relativistic particle, 
\begin{equation}
\varepsilon^2 = (mc^2)^2 + (p_xc)^2+ (p_yc)^2+ (p_zc)^2.
\end{equation}

Thus, the number of states (discrete points) in a sphere of radius $\varepsilon$ corresponds to the number of states inside an ellipsoid with volume $4/3\pi p_xp_yp_z$. The non-zero mass only shifts the ellipsoid's centre without affecting its volume. To determine a mode's space with positive numbers $l_i$ and momentums $p_i$, only one octant should be considered, so the number of modes can be evaluated as:
\begin{equation}
dN=1/8 dl_xdl_ydl_z=\frac{L_x L_y L_z}{h^3}dp_xdp_ydp_z=\frac{V}{h^3}4\pi p^2 dp.
\end{equation}

Thus, the mode density becomes:
\begin{equation}
d\rho=\frac{4\pi p^2 dp}{h^3}.
\end{equation}

This is valid for  massive and massless particles in a general (relativistic) case even for an infinite box size. In the case of $L\to\infty$, the mode distribution becomes continuous.

This result is also valid if a boundary condition $\psi(0)=\psi(L)=0$ of an isolated box is relaxed to a continuity condition $\psi(0)=\psi(L)$. In  such a condition, negative values of $l_i$ are allowed, and $\vec{p}$  becomes a vector with values in a full space (and not just one octant). However, only wavelength multiplets  being equal to the size of the box are now allowed (opposite to the half-wavelength multiplets). This gives a $(1/2)^3$ reduction in the number of modes that is numerically equal to one octant condition.

\subsection{Mode Energy}
The electromagnetic field  can be quantized starting from Maxwell's equations and performing Fourier analysis for modes in the box with periodic conditions~\cite{MandlBook}. For such systems, the Hamiltonian becomes equivalent to an infinite set of oscillators. 
The mode energy {levels {are,}  
as for a harmonic oscillator}: 
\begin{equation}
E=\hbar\omega\left(\frac{1}{2}+n\right),
\label{eq:mode_energy}
\end{equation}
where $n$ is the number of quanta with energy $\hbar\omega$---the smallest value of energy that can be taken or added to the mode with $\omega$ frequency. For the electromagnetic field, a quantum is associated with a single photon; thus, $\varepsilon=\hbar\omega=pc$.

\subsection{Statistics}
For this part, it is worth mentioning that conventionally and historically, black-body radiation is described for a closed box in thermodynamical equilibrium 
with the temperature $T$ and the volume $V$ being fixed. In statistical mechanics, this is, however, ambiguous since it may correspond to any of the following. 
\begin{itemize}
\item The grand canonical ensemble ($T,V,\upmu$ fixed), where $\upmu$ is the chemical potential: 
The system can exchange energy and particles with a reservoir, so that various possible states of the system can differ in both their total energy and the total number of particles.
\item The canonical ensemble ($T,V,N$ fixed): The system can exchange energy with the heat bath, so that the states of the system will differ in total energy; the number of particles is fixed.
\end{itemize}

In the equilibrium, the probability of a state with energy distribution $E_i$  is described by the following relations. For canonical ensemble, 
\begin{equation}
\mathrm{p}_i=\frac{e^{-E_i/(kT)}}{Z},
\end{equation}
where $Z$ is the partition function assuring probability normalisation, 
\begin{equation}
Z=\Sigma_i e^{-E_i/(kT)}, 
\end{equation}
 and $k$ is the Boltzmann constant.

The most powerful and general way to obtain this probability is using the information-theoretic Jaynesian maximum entropy approach {\cite{Jaynes1,Jaynes2}}.  

Similarly, for the grand canonical ensemble:
\begin{equation}
\mathrm{p}_i=\frac{e^{(N_i\upmu-E_i)/(kT)}}{Z},
\end{equation}
\begin{equation}
Z=\Sigma_i e^{(N_i\upmu-E_i)/(kT)}.
\end{equation}

For the black-body radiation, the number of photons is not conserved, so it is considered that the grand canonical ensemble is the right choice, although even in some recent studies the opposite choice is made (see critiques in~Ref. 
\cite{Mishra2017}, for example). The confusion is supported by the feature that for black-body radiation, $\upmu=0$ is assumed, so both probabilities become identical. The choice of  $\upmu=0$ 
is a consequence of the property that the black-body radiation in a closed cell should be completely defined  by only two macroscopic parameters: $T$ and $V$. As a result, the system pressure is defined only by temperature, and the Gibbs free energy 
 $G=\upmu N$ 
becomes 0 (actually, not well-defined), so for a system with a non-zero $N$, it is required to have $\upmu
 =0$~\cite{Kelly1981}.

In a case of non-interacting bosons, each available single-particle level (mode) forms a separate thermodynamic system in contact with the reservoir. Thus, the analysis of the system behaviour can be conducted within a single mode, and then the properties of the whole system can be obtained by integrating it over the modes with their densities. For a single mode, the grand canonical partition function becomes (omitting {$\upmu$} term):
\begin{equation}
Z=\Sigma_{n=0}^\infty e^{-\hbar\omega(n+1/2)/(kT)}=e^{-\hbar\omega/(2kT)} \frac{1}{1-e^{-\hbar\omega/(kT)}}.
\end{equation} 

And each state probability within a mode is:
\begin{equation}
\mathrm{p}_i=\mathrm{p}(n)=\frac{e^{-\hbar\omega(n+1/2)/(kT)}}{Z}=
e^{-\hbar\omega n/(kT)}(1-e^{-\hbar\omega/(kT)}).
\end{equation} 

Interestingly, the probability {does not contain the zero-point energy term}, i.e., the probability would be identical if this term {was} omitted from the beginning.
For the average energy evaluation, it is convenient to use the following property of the partition function:
\begin{equation}
\langle E\rangle=\frac{\Sigma_{n=0}^\infty E_n e^{-E_n \beta}}{Z}=\frac{dZ}{d\beta} \frac{1}{Z} = - 
\frac{d({\rm log}Z)}{d\beta}=\hbar\omega\left(\frac{1}{2}+\frac{1}{e^{\hbar\omega/(kT)}-1}\right),
\label{eq:e_av1}
\end{equation}
where $\beta = 1/(kT)$ is introduced for simplicity.
The average number of particles can be obtained from the average property:
\begin{equation}
\langle E \rangle=\langle
\hbar\omega\left(\frac{1}{2} +  n\right)\rangle=\hbar\omega\left(\frac{1}{2}+\langle n\rangle \right),
\label{eq:e_av2}
\end{equation}
and by comparing Equations (\ref{eq:e_av1}) and (\ref{eq:e_av2}), it can be expressed as follows:
\begin{equation}
\langle n\rangle=\frac{1}{e^{\hbar\omega/(kT)}-1}.
\end{equation}

For the canonical ensemble of bosons, the number of particles, $N$, is fixed (the system cannot exchange particles with a reservoir) and the modes cannot be considered as independent thermodynamical systems. In the limit of $n_i\to\infty$ (the number of particles in each mode is extremely large), one can, however, obtain a surprisingly similar average number of particles for each mode (energy level) as in the case of the canonical 
ensemble \cite{BEcan}. 

\subsection{Degeneracy}
In the above considerations, the feature that each mode for the electromagnetic field has a degeneracy $g=2$ corresponding to the two polarisations was omitted. This, effectively, makes the mode 
density twice as high.

\subsection{Energy Density}
Combining the results from Section \ref{sec:density}, one obtains the Planck's law:
\begin{equation}
w(p)dp=2 \frac{4\pi p^2 dp}{h^3} \varepsilon \left( \frac{1}{2}+\frac{1}{e^{\varepsilon/(kT)}-1}\right),
\end{equation}
from where the vacuum density can be deduced:
\begin{equation}
\rho(p)dp=\frac{4\pi p^2 dp}{h^3}.
\label{eq:vac_density}
\end{equation}

Following the derivation, let us summarise some features.
\begin{itemize}
\item The vacuum density is numerically equal to the density of the modes---its  $1/2$ factor from the zero-level is compensated by the degeneracy $g=2$;
\item Modes and their zero-point energies appear for the electromagnetic field after  Fourier transformation, so, strictly speaking, assuming  a mode  occupies some volume (even on average) is improper;
\item Derivation of the statistical distribution for modes assumes they are independent and that the number of quanta in each mode is high: this may leave some speculation whether the vacuum density is described by  Equation~(\ref{eq:vac_density}) in cases when there are few or no quanta (associated with real photons);
\item The vacuum density is temperature- and energy-independent (non-thermal, linear); in current derivation, this appears as a consequence of 
several features: modes are independent and each mode's microstate energy has a zero-level energy that is proportional to the mode quantum's energy with the same factor for each mode;
\item The vacuum density and its energy density are infinite if there is no cutoff at high energies/frequencies; the introduction of such an ultraviolet cutoff is a viable option, for example, in doubly special relativity~\cite{Mishra2017}.
\end{itemize}

Similarly to photons, one may apply the discussed derivation to fermions. The mode density, as discussed in {Section} \ref{s:density_of_modes} 
above, is the same for massive and massless particles. The degeneracy for  fermions is $g=2$. Quark fermions have additional factor 3 degeneracy due to their colour. The vacuum density should also follow Equation~(\ref{eq:vac_density}) since Bose--Enstein or Fermi--Dirac statistical terms are not relevant. The quantum energy of each mode is actually $((mc^2)^2+(pc)^2)^{1/2}$, where $m$ is the mass of a fermion. Assuming now that virtual fluctuations should appear as fermion--antifermion pairs with opposite spin and momentum, their density should also be described by the same equation, in which $p$ always refers to the momentum of a single fermion or antifermion. Finally, let us note that for boson pairs ($W^+W^-$) the vacuum density expression is also the same as for fermions since the statistical term is relevant only for real quanta; however, degeneracy $g=3$ for a spin of 1 should be used. 

\section{The Vacuum Permeability and Permittivity}
\label{sec:MuEpsi}
\subsection{Calculation with ZPF Density Using Oscillator Model}
\label{sec:MuEpsi_calc}
In order to perform vacuum permittivity calculations, we consider some of the assumptions of Ref.~\cite{Leuchs2023}, in particular:
\begin{itemize}
\item each virtual fermion--antifermion pair behaves as a harmonic oscillator with the levels separated by $\hbar\omega$; specifically, this is the energy gap between the ground state of a virtual pair and the excited state where both particles become real.
 \end{itemize}
 This assumption is compatible with the mode energy~(\ref{eq:mode_energy}) of a quantised electromagnetic field. We further try to make the calculation 
in Ref.~\cite{Leuchs2023} more consistent with modes and their quanta. This brings the following assumptions:
 \begin{itemize}
\item a virtual fermion--antifermion pair becomes real if $\varepsilon=2((mc^2)^2+(pc)^2)^{1/2}$ is added to the mode so the oscillator frequency is actually $\hbar\omega=\varepsilon(p)$ instead of 
 $2 mc^2$;
\item we use the vacuum (mode) density for the estimation of $\epsilon_0$  instead of the average volume following the alternative proposed at the end of~\cite{Leuchs2023}.
\end{itemize}

Following the study in Ref.~\cite{Leuchs2023}, if one arranges the two-level approximation, so that only transitions between the ground and the first excited states of the oscillator are relevant, one finds that the maximum possible induced dipole moment becomes:
\begin{multline}
\mathrm{d}_\mathrm{max}= q_e\langle \psi_1 | \hat{x} | \psi_0 \rangle = q_e \int_{-\infty}^{\infty}\left( \frac{m\omega}{\pi\hbar} \right) ^{1/4} e^{-\frac{m\omega x^2}{2\hbar}}   
 \left( \frac{m\omega}{\pi\hbar} \right) ^{1/4} e^{-\frac{m\omega x^2}{2\hbar}} \sqrt{2} \left( \frac{m\omega}{\hbar} \right) ^{1/2}x dx \\
= q_e \int_{-\infty}^{\infty} \sqrt{\frac{2}{\pi}}  \left( \frac{m\omega}{\hbar} \right)  e^{-\frac{m\omega x^2}{\hbar}} x^2 = q_e \sqrt{\frac{2}{\pi}} 
\sqrt{\frac{\hbar}{m\omega}} \int_{-\infty}^{\infty} e^{-z^2} z^2 dz = q_e\sqrt{\frac{\hbar}{2m\omega}}.
\end{multline}
   Here, $q_e$ denotes the electron charge.

The time-averaged induced dipole moments following calculations in Ref.~\cite{Leuchs2023} 
(Equation (A3)) can be expressed as follows:
\begin{equation}
\mathrm{d}=\frac{2\mathrm{d}_\mathrm{max}^2}{\hbar\omega}E=\frac{q_e^2}{m\omega^2}E.
\label{eq:p}
\end{equation}

The expressions for $\mathrm{d}_\mathrm{max}$ and $\mathrm{d}$ here differ from Equations (A2) and (A3) of Ref.~\cite{Leuchs2023} by factors $1/\sqrt{2}$ and $1/2$, respectively, as soon as
$\hbar\omega = 2 mc^2$ is assumed. Equation~(\ref{eq:p}) is same as Equation (25) obtained in Ref.~\cite{Margan2011} within a semiclassical treatment. 

The vacuum permittivity calculation then can be estimated as follows, while assuming only one type of fermion so far: 

\begin{equation}
\epsilon_0=\int_0^{p_\mathrm{max}} \frac{4\pi p^2\,dp}{h^3} \frac{q_e^2}{m\omega^2} = \frac{q_e^2}{4\pi\hbar c}\frac{1}{2\pi} \frac{1}{mc^2} \int_0^A \frac{(pc)^2 d(pc)}{(pc)^2+(mc^2)^2},
\label{eq:epsi0density}
\end{equation}
where the cutoff on $pc$ at $A$ is introduced in order to have a finite integral. The introduction of the cutoff is just instrumental here, and its {possible physical origin is discussed below in
this Section} and in Section \ref{sec:conclusions}.
 {One} can rewrite the above equation by using fine-structure constant, {$\alpha$:} 
 \begin{equation}
\frac{1}{\alpha}=\frac{1}{2\pi} \frac{1}{mc^2} \int_0^A \frac{(pc)^2 d(pc)}{(pc)^2+(mc^2)^2}=
  \frac{1}{2\pi} 
 \left[
 \frac{A}{mc^2} - \tan^{-1}\left(\frac{A}{mc^2}\right)  
\right]. 
\end{equation}

One can see that in order to obtain $1/\alpha\approx137$, a value of $A$ should be on the order of $A\approx(2 \pi / \alpha) mc^2 \approx 861 mc^2$, where the 
factor $2\pi/\alpha$ resembles the inverted QED correction needed for the electron magnetic moment (one-loop result for the anomalous magnetic moment). If all known charged elementary particles are considered, the $\epsilon_0$ and $1/\alpha$ expressions contain a sum over charged fermions, or more generally, over all charged elementary particles:

\begin{equation}
\frac{1}{\alpha}= \frac{1}{2\pi} \sum_i Q_i^2 c_i \frac{g_i}{2} \left[\frac{A}{m_i c^2} - \tan^{-1}\left(\frac{A}{m_i c^2}\right)  \right],
\label{eq:1_alpha}
\end{equation}
where charge scale $Q_i=1$ for leptons and $W$-bosons, $Q_i=-1/3$ for $d, s, b$ quarks, and $Q_i=2/3$ for $u, c, t$ quarks; the degeneracy, $g_i$,
due to a spin is $2$ for fermions and $3$ for $W^-W^+$ pairs, while the degeneracy, $c_i$, due to a colour is 3 for quarks and 1 for the other types. If one keeps the upper energy value the same for each type, a value of {the threshold} $A\approx292$~MeV is needed to reach the measured value of $\epsilon_0$ or $\alpha$. The most important contribution is coming from electrons and the second contribution from $u$ quarks, for which $m_u=1.5$~MeV was used. Other contributions are at the percent level and below. The threshold, $A$, is much higher than $m_e c^2$, and the greatest contribution is coming from the ultra-relativistic pairs since, for $A=m_e c^2$, only $0.1\%\epsilon_0$ is reached. This poses  doubt if the used non-relativistic harmonic oscillator model is reasonable. 
One possibility to lower the threshold, $A$, and to avoid relativistic pairs is to assume that instead of the real charges, one should use bigger, unscreened charges.

The obtained value of $A$ is above or on a level of the chiral symmetry breaking value ($\approx$100~MeV or pion mass), the energy above which the quark condensate should disappear~\cite{Rugh2000}. One should investigate how to incorporate this effect. A possible solution could be to consider that for the quarks the $A$ threshold is set to this level. Since the quark contribution to $1/\alpha$ is sub-dominant, this should not strongly affect the evaluation 
in Ref.~(\ref{eq:1_alpha}). The current threshold is three orders of magnitude below the electroweak unification (246~GeV). Such a threshold could be a physically motivated choice for the leptonic virtual pairs; however, $\epsilon_0$ would not be matched with the measured value.

It is possible to make the above calculations with vacuum density consistent with the calculations in 
Ref.~\cite{Leuchs2023},  
which exploit the concept of the average volume occupied by virtual pairs. In this paper,
we can numerically estimate the average virtual pair volume as the inverted density. In order to make both calculations compatible, one needs to assume that the separation of energy levels  is energy independent, and it is equal to 
$2 mc^2$, 
as in Ref.~\cite{Leuchs2023}. This allows one to split  the constant multiplication in Equation~(\ref{eq:epsi0density}) from the integral, which becomes just the {vacuum density,} $\rho=\int_0^{A/c} 4\pi p^2 dp / h^3$. In order the average volume to be proportional to $\lambda_C^3=(\hbar/(mc))^3$, one needs to set the limit, $A$, proportional to $mc^2$: $A=amc^2$, i.e., to assume a different threshold for each elementary particle type. 

With $A=amc^2$ thresholds and $\hbar\omega = 2 mc^2$, the expression for $1/\alpha$ reads:
\begin{equation}
\frac{1}{\alpha}=  \frac{1}{2\pi}  \sum_i Q_i^2 c_i \frac{g_i}{2} \frac{1}{(mc^2)^3} \int_0^{amc^2} (pc)2d(pc) =\frac{1}{2\pi}  \sum_i Q_i^2 c_i 
\frac{g_i}{2}  \frac{a^3}{3}.
\end{equation}

With this threshold, one obtains a factor $a= \sqrt[3]{6\pi / (\sum_i Q_i^2 c_i (g_i/2) \alpha)}\approx6.5$ ({we consider $g_i=3$ for $W^+W^-$, so 
$\sum_i Q_i c_i (g_i/2) ^2=9.5$, and not 9 as in Ref.~\cite{Leuchs2023}}) 
and the average volume reads:
\begin{equation}
\langle V \rangle = 1/n = 1 / \int_0^{A/c} \frac{4\pi p^2 dp}{h^3}  =
\frac{6\pi^2}{a^3} \left(\frac{\hbar}{mc}\right)^3 \approx 0.22 \left( \frac{\hbar}{mc} \right) ^3 ,
\end{equation}
 what is consistent with the calculations in Ref.~\cite{Leuchs2023}, considering the  difference in the dipole moment by a factor of 2 and the differences in $g_i$ 
for $W$-bosons. 

\subsection{Relations between Magnetic and Electric Dipole Moments of Virtual Pairs}
Here, we reiterate to avoid the oscillator approximation approach and to start from a magnetic moment of each fermion in the pair along with an electric dipole moment of a pair in order to arrive at the evaluation of  $\epsilon_0$ and $\mu_0$. We provide below an ansatz for calculating magnetic and electric dipole moments  for virtual pairs 
that keeps the $c,\, \epsilon_0,$ and $\mu_0$ interrelation.

First, it is worth mentioning that in some models of vacuum description~\cite{RefUrban} and for  real fermion {gas} 
(Equations ({59.4}) and ({59.12}) {in} Ref. \cite{LL}), there are the following dependences for permittivity and permeability:
\begin{equation}
1/\mu_0=\sum_i f(\beta_i^2),
\label{eq33}
\end{equation}

\begin{equation}
\epsilon_0=\sum_i f((\mathrm{d}_i/2)^2),
\label{eq34}
\end{equation}
where $\beta$ is a magnetic moment of a single fermion, $\mathrm{d}$ {is an electric dipole moment} of a fermion--antifermion pair, and function $f$ is of the same dependence in Equations (\ref{eq33}) and (\ref{eq34}) as soon as the corresponding potentials are of the same expression, namely, $U=-\vec{\mathrm{d}}\vec E$ and $U=-\vec\beta\vec B$. 
Given $c=1/\sqrt{\epsilon_0\mu_0}$, one finds that this leads to the following relationship between $\beta$ and $\mathrm{d}$:
\begin{equation}
\beta=(\mathrm{d}/2)c.
\label{eq:betaomega}
\end{equation}

The magnetic moment can be obtained by comparing relativistic energy for an electron in a magnetic field~\cite{RelMM} \mbox{(in SI units)}:
\begin{equation}
E_M=\sqrt{m^2c^4+p_z^2c^2 + 2q_e\hbar c^2 B(n+1/2 -gs/2)} \\
\approx \varepsilon_f + \frac{q_e\hbar c^2 B}{2\varepsilon_f} (2n+1-gs),
\label{eq:relenergyB}
\end{equation}
where we consider the first order approximation as valid for common fields: $\beta_B B \ll m_ec^2$ (here, $\beta_B$ is the Bohr magneton).

This approximation can be compared to the non-relativistic energy levels in~ Ref.\cite{LL} (\mbox{in SI units}): 
\begin{equation}
E_M^{NR}=\frac{p_z^2}{2m}+\frac{q_e\hbar B}{2m}(2n+1-gs).
\end{equation}

Comparing relativistic and non-relativistic expressions, one can see that they match if 
$mc^2$ 
is substituted with fermion energy, $\varepsilon_f$. Thus, the following relativistic magnetic moment can be used instead of the Bohr magneton, $\beta_B = q_e\hbar/(2m)$:
\begin{equation}
\beta=\frac{q_e\hbar}{2\varepsilon_f/c^2}.
\label{eq:relMM}
\end{equation}
For electric dipole moments, one can use the following classical expression:
\begin{equation}
\mathrm{d}=Q q_e x,
\label{eq:omega}
\end{equation}
where $x$ is considered a 
``distance'' between fermions.

By comparing expressions (\ref{eq:relMM}) and (\ref{eq:omega}), one can realise that Equation~(\ref{eq:betaomega}) is satisfied if: 

\begin{itemize}
\item the magnetic moment, $\beta$, for fermions with charge $Q q_e$ has an additional factor $Q$;
\item for electric dipole moment, $\mathrm{d}$, the ``distance'', $x$, is defined as $x=\hbar/(\varepsilon_f/c)$.
\end{itemize}

If the last point is seen as the Heisenberg uncertainty, this requires that the fermions travel at a speed of light or they are of zero mass.

\subsection{Other Considerations for the Virtual Pair Models}

If a virtual electron--positron pair is seen as a positronium~\cite{Mainland2020} it is worth establishing a connection with an up-to-date description of the positronium states. In the classical description of such a system, similarly to the electron in hydrogen, it is assumed that the electron and positron are orbiting around the centre of mass. Thus, the energy levels are quantised as in the Bohr model but using a reduced mass equal to $m_e/2$. The precise calculation of the bound states comes from the Bethe--Salpeter equation. Noticeably, there are two solutions with the orthogonal states: one in which particles can be bound at atomic-like distances, similarly to the Bohr model, and the other with the nuclear-like quantised distances~\cite{Patterson2023}. The latter states have zero energy, and thus they could be very promising for describing virtual pairs.  

It is worth mentioning that the optical properties, namely, polarisation and nonquantum entanglement~\cite{Simon2010}, can be described using the mechanical model analogy~\cite{Qian2023}. This mechanical model for a  photon or a 2-dimensional (2D)  beam consists of two masses, each representing the eigenvalue of a polarisation coherence matrix. The polarisation and entanglement are then quantitatively associated with mechanical concepts of the centre of mass and the moment of inertia through the Huygens--Steiner theorem for rigid body rotation. Although it 
may be a mere coincidence of this analogy, one may want to search for a deeper physical meaning. 

\section{Photon Propagation and Propagation Time Dispersion}
\label{sec:prop}

Most theories explaining electromagnetic vacuum properties with virtual fluctuations assume that particle--antiparticle pairs are continuously appearing in the vacuum and their lifetimes are limited by the Heisenberg uncertainty. We detail here several theories and the lifetimes assumed:
\begin{itemize}
\item $\tau\approx\hbar/(2 mc^2$), where this time, in particular, serves to define the size/volume of the virtual pair~\cite{Leuchs2023};
\item $\tau=\hbar/(K\times 2mc^2)$ with the best fit for $K\approx31.9$, where the lifetime modified in the presence of  electromagnetic field serves to evaluate $\epsilon_0$ and $\mu_0$~\cite{RefUrban};
\item $\tau=\hbar/(\alpha^5mc^2)$ since, after interaction with the photon, the virtual pair forms a quasi-stationary state in Ref.~\cite{Mainland2020}.
\end{itemize}

In a bare vacuum, the dielectric and magnetic moments of the virtual fermion pairs are absent, and thus $\epsilon_0$ and  {$1/\mu_0$} become 0. 
The speed of light in the Maxwell equations in a vacuum is 
 {$c=\left[(1/\mu_0)/\epsilon_0\right]^{1/2}$]} 
and becomes not well defined~\cite{Leuchs2020}. This indicates that the photon propagation is tightly connected with the presence of virtual fluctuations. It is 
 quite natural then, to assume that the photon propagation speed is actually finite only because the photon is delayed at each interaction with virtual fluctuations by their annihilation time, i.e., there is no additional propagation delay in the bare vacuum. By knowing the average lifetime of a virtual pair, $\tau$, and the total time $T=L/c$ needed to cover a distance $L$, one can straightforwardly estimate the total number of interactions, $N=T/\tau$, and its fluctuation, $N^{1/2}$, which gives an approximate fluctuation time estimate:

\begin{equation}
\sigma_T= N^{1/2}\tau = \sqrt{\frac{\tau}{c}} \sqrt{L}\, .
\end{equation}

Thus, for the three theories mentioned above, this fluctuation time becomes:
\begin{itemize}
\item $\sigma_T\approx1.5\,\mathrm{fs}\sqrt{L\mathrm{[m]}}$ for consideration in Ref.~\cite{Leuchs2023};
\item $\sigma_T\approx0.26\,\mathrm{fs}\sqrt{L\mathrm{[m]}}$ for {for consideration in Ref.}~\cite{RefUrban} (this can be compared with a more precise estimation at $\sigma_T\approx0.05\,\mathrm{fs}\sqrt{L\mathrm{[m]}}$ given in
 Ref.~\cite{RefUrban});
\item $\sigma_T\approx0.46\,\mathrm{ns}\sqrt{L\mathrm{[m]}}$ for {for consideration in Ref.}~\cite{Mainland2020}.
\end{itemize}

Let us note that the above calculations are meant to provide an order of magnitude. For a better {estimate}, one may assume that the photon is delayed at each interaction only by a portion of a lifetime since the photon interaction can occur at a random moment of the pair's appearance from the vacuum, and one can consider a smooth distribution for the possible 
{lifetimes.} The fluctuation in propagation time  can actually be removed completely if one assumes that during each interaction with time $\tau_i$, a photon propagates for a distance of $c\tau_i$, which may look less intuitive. Indeed, one would need to explain the photon energy transfer in space inside a virtual pair that absorbs the photon.

Nowadays, the strongest constraints on the photon propagation time fluctuation are established by astrophysical observations, mainly 
GRBs (gamma ray bursts) and pulsars~\cite{Urban8,Urban9}. The current limits are at $0.2-0.3\,\mathrm{fs\,m}^{-1/2}$. This means that if the estimate presented here is correct, at least at the order of magnitude, the model in Ref.~\cite{Mainland2020} to be excluded, while the other two considered  models are still viable.
 
The dependence of the fluctuations as $L^{1/2}$ shows that for the measurement, the time resolution has a stronger impact compared to the photon path length. Actually, the astrophysical measurements are based on the  observation of events with a duration of the order of $10^{-3}\,\mathrm{s}$ at distances of megaparsecs, 
and these measurements are hardly improvable due to the finite  size of the Universe and the intrinsic event durations. Instead, the current state of 
the art of laser technologies allows the generation of femtosecond and even attosecond light pulses, and a pulse duration evaluation is conducted using the autocorrelation, i.e., {probing} the 
{pulse by its own copy,}
so the precision is better than the pulse width.

An experiment able to reach the required sensitivity can be realised with femtosecond laser pulses propagating over a few kilometres in a vacuum tunnel without reflections. In the end of the tunnel, the beam is measured with the autocorrelation system. A minimal realisation of the autocorrelation system can consist of a beam splitter, beam routing with three unprotected, gold, single-layer mirrors per  split beam and a second harmonic crystal. The mirror system generates a variable delay between the two beam parts. Using a commercially available laser source with 
 {$\mathcal{T}=2$}\,fs pulse width ($\sim$5 fs 
full-width half-maximum, FWHM), the vacuum optical path of L = 10\,km ({such tunnels are routinely used in modern gravitational wave detectors}) and a pulse duration measurement device with 
 {$\sigma_\mathcal{T}=1\%$} precision, one could already set a limit that is two orders of magnitude lower  compared to the existing ones, as  is estimated below:
\begin{equation}
\sqrt{\mathcal{T}^2+\sigma^2 L}-\mathcal{T}=\sigma_\mathcal{T}\mathcal{T},
\end{equation}

\begin{equation}
\sigma=\mathcal{T}\sqrt{\frac{\sigma_\mathcal{T}(2+\sigma_\mathcal{T})}{L}}\approx2\sqrt{\frac{0.02}{10^4}}\,\mathrm{fs\,m^{-1/2}}\approx0.003\, \mathrm{fs\,m^{-1/2}}.
\end{equation}

If one considers the laser with an attosecond pulse duration, the generated pulse would be broadened up to a 16-as duration  after only a  1-cm propagation with $\sigma_T\approx0.05\,\mathrm{fs}\sqrt{L\mathrm{[m]}}$. Interestingly, the world's shortest pulse generated is measured at 
{the FWHM of $43\pm1$~as} \cite{Gaumnitz2017}. Such a pulse already after 1\,cm would be at {the  FWHM of 57~as} or, alternatively, a pulse with a negligible duration would have a {duration of 43~as } after 13\,cm. If fluctuations at such scales are real, currently measured attosecond pulses could be already close to the limit of the generation of shortest pulses and the measurement of their duration. 
Setting limits to the fluctuations in the speed of light with the best available attosecond laser and a 1--50\,cm path can be a viable alternative for such measurement. 
Finally, let us note that since the fluctuation in propagation time  is frequency-independent, the frequency-resolved optical gating pulse characterisation in the time and frequency domains should provide robust effect measurements and enable its discrimination from the matter effects. This effect  also has distinct dependence from the distance, so performing additional measurements with a shorter path is another way to control the systematic effects. 
\section{Conclusions and Discussions}
\label{sec:conclusions}
There is certainly an interest in the physics community concerning the idea of virtual fluctuations being at the origin of vacuum electromagnetic properties. As it is shown, similarly to $\vec{P}$, polarisation of the medium, $\epsilon_0 \vec E$, can be associated with the polarisation of the vacuum, and similarly, {$1/\mu_0 \vec{B}$} can be associated with  vacuum magnetisation. This view can clarify the historical controversy between $\vec{H}$ versus $\vec{B}$ and $\vec{D}$ versus $\vec{E}$~\cite{Roche2000}. In the absence of matter, $\vec{H}$ becomes the response of the vacuum to the external field $\vec{B}$, while in the presence of matter, it is a combination of the vacuum response to the external field and the matter's magnetisation (permanent, or induced by the field $\vec{B})$. This follows Faraday and Maxwell, which makes $\vec{H}$ the cause of $\vec{B}$ and similarly for $\vec{D}$ and $\vec{E}$.

The authors of~\cite{Leuchs2023} show how one can evaluate vacuum polarisation, namely $\epsilon_0$ and $1/\mu_0$,  
within QED from the  interactions of photons with electron pairs using the Feyman diagrams description. In this approach, one should deal with a high momentum cutoff, as this prevents the exact calculation of $\epsilon_0$ without the introduction of a new variable.

A tempting approach to avoid infinite integrals that require cutoffs or renormalisations is to use a constant or an average volume occupied by virtual pairs along with average magnetic and electric moments. Here, 
we revisit the mathematical origin of zero-point energy appearing in Planck's law derivation in order to stress the feature that this energy appears in the mode 
space  associated with particle momentum. Actually, it is impossible to estimate the volume occupied by a virtual pair with a known momentum due to the Heisenberg uncertainty. The evaluation of $\epsilon_0$ and $\mu_0$  
can be performed using vacuum density in the form $\rho(p)=4\pi p^2/h^3$ instead of using an averaged pair volume. Moreover, from the mathematical expression of  Plank's law, each mode has its own frequency, $\omega$, or energy quanta, $\hbar\omega$; thus, it is more logical to assume that the zero-point energy, $\frac{1}{2}\hbar \omega$, should  
generate virtual pairs of fermions with non-zero momentum per fermion. Here, we show how these calculations can be performed starting from the 
ideas in Ref.~\cite{Leuchs2023}. 
In order to obtain a finite value for $\epsilon_0$, a threshold on the maximum fermion momentum is needed. The value of this threshold in our calculation does not provide any intriguing coincidence; however, the calculation shows that most of the fermions should be relativistic. Thus, the use of a non-relativistic quantum harmonic oscillator as a model could be argued. One could use, however, bigger, unscreened charges in order to lower this threshold. We do not propose any physical explanation of the obtained threshold at 292~MeV. A physically motivated choice for such threshold for quarks could be a chiral symmetry breaking energy ($\approx$100~MeV), while for leptons, it could be the electroweak unification energy ($\approx$256~GeV). This 
indicates an incompleteness of our model. An understanding of how to incorporate the breaking of symmetry and the generation of mass, i.e., the Higgs mechanism, could be the key to explaining the threshold. We note here for completeness that calculating $\epsilon_0$  within QFT requires  knowledge of the Landau pole energy. Its value, obtained in~\cite{Leuchs2023}, does not provide any intriguing coincidence either.

As it is shown, dealing with infinite integrals in the calculations involving virtual pair density takes two distinct paths. In one approach, 
explored in
 Ref.~\cite{Leuchs2023}, 
it is considered that the volume occupied by virtual pairs has a finite, momentum-independent value, and the vacuum properties are evaluated using this volume scale as a parameter. In~\cite{Leuchs2023}, it is demonstrated that this approach is consistent with a calculation within QFT, which, in turn, requires the Landau pole energy as a parameter. In another approach, in this study, the density of virtual pairs is used, following its zero-point field origin. The calculations require a cutoff at a high momentum value, which becomes a parameter. The average volume per pair  can still be estimated as the inverted density. In a particular case of momentum-independent transition between the virtual fluctuation states, the calculation with the average volume and the density become equivalent. The predictive power of both approaches should be quantified by the number of properties that can be estimated based on such parameters (pair volume or momentum cutoff) and a consistency with the existing observations. The properties may include the vacuum permeability, the vacuum density, and the cosmological vacuum energy density. Meanwhile, the observations should include quantum phenomena such as the interference patterns in double slit experiments and the stability of atoms and their energy levels. In particular, the presence of  virtual fluctuations in the vacuum must have an impact on the cosmological scales. 

The introduction of the negative gravitational mass for the antiparticles makes each massive virtual pair a gravitational dipole. This model solves several cosmological issues at once, including effects ascribed to  dark matter and  the cosmological constant problem~\cite{Hajdukovic2020}. The recent results from the ALPHA-g experiment on antihydrogen atoms show that antimatter accelerates towards the Earth with $0.75g\pm0.13$ g (statistical+systematic) $\pm0.16$ g (simulation)~\cite{ALPHAg2023}, so the theories involving antigravity for antimatter are \mbox{hardly viable}.

Several theories that describe photon interactions with virtual pairs include the pair lifetime~\cite{Leuchs2023, Mainland2020, RefUrban}. We speculate that one of the consequences of this assumption is that at each interaction, the photon is trapped until the pair is annihilated, and this may give rise to a fluctuation in the photon propagation time. This fluctuation can be estimated for the mentioned theories. In particular, 
the theory in
Ref.~\cite{Mainland2020} 
could already be in conflict with the available measurements. For the fluctuations predicted for theories in Ref.~\cite{Leuchs2023,RefUrban}, 
one needs measurements to be made with available ultrafast laser technologies. The observation of the fluctuation in photon propagation time  would certainly be incompatible with quantum field theory.

\vspace{6pt} 

\authorcontributions{Conceptualization, C.H. and V.K.; methodology, C.H. and V.K.; formal analysis, V.K.; writing---original draft preparation, V.K.; writing---review and editing, C.H. and V. K.; funding acquisition, V.K. All authors have read and agreed to the published version of the manuscript.}

\funding{The APC and the study was funded by the Istituto Nazionale di Fisica Nucleare, Italy (INFN), including the grant for training activities 195393 attributed to the project ReWOLF-Cub, Italy.}

\dataavailability{Calculation notebook is published as} \url{https://doi.org/10.5281/zenodo.10457877} 

\conflictsofinterest{The authors declare no conflict of interest.} 

\abbreviations{Abbreviations}{
The following abbreviations are used in this manuscript:\\

\noindent 
\begin{tabular}{@{}ll}
GRB & Gamma-ray burst\\
FWHM & {Full-width half-maximum}\\
SED & Stochastic electrodynamics\\
QED & Quantum electrodynamics\\
QFT & Quantum field theory\\
QM & Quantum mechanics\\
ZPF & Zero-point fluctuations
\end{tabular}
}

\begin{adjustwidth}{-\extralength}{0cm}

\reftitle{References}

\PublishersNote{}
\end{adjustwidth}
\end{document}